\documentclass[letterpaper]{aux/easychair}

\def\final{}      

\ifdefined\final \else
  \def\anon{}
\fi
\usepackage[T1]{fontenc}
\usepackage{graphicx}
\usepackage{textcomp}
\usepackage{xcolor}
\usepackage{multirow}
\usepackage{threeparttable}
\usepackage[htt]{hyphenat}
\usepackage{fancyhdr}

\newcommand{\mytitle}{On the Effectiveness of Instruction-Tuning Local LLMs \\for
Identifying Software Vulnerabilities}
\newcommand{\myauthors}{Park et al.}

\usepackage{etoolbox}
\AtBeginEnvironment{tablenotes}{\footnotesize}

\newcommand{\svi}{Software Vulnerability Identification}
\newcommand{\svifull}{\svi \ (SVI)}
\newcommand{\svd}{Software Vulnerability Detection}
\newcommand{\svdfull}{\svd \ (SVD)}

\newcommand{\codetfiveex}{{\texttt{codet5-large}}}
\newcommand{\codebertex}{\texttt{codebert-base}}
\newcommand{\gptthreeex}{\texttt{gpt-3.5-turbo}}
\newcommand{\gptfourex}{\texttt{gpt-4-0613}}
\newcommand{\codellamaex}{\texttt{codellama-7b-instruct}}
\newcommand{\llamathreeex}{\texttt{llama3.1-8b-instruct}}

\makeatletter
\def\thm@space@setup{\thm@preskip=1pt
\thm@postskip=1pt}
\makeatother

\newcommand{\code}[1]{\lstinline!#1!}

\ifdefined\acm
  
\fi

\DeclareMathAlphabet{\mathcal}{OMS}{cmsy}{m}{n}

\newcommand{\keywords}[1]{\par\smallskip\noindent\textbf{Keywords:} #1\par\smallskip}

\begin{document}

\title{On the Effectiveness of Instruction-Tuning Local LLMs for
Identifying Software Vulnerabilities\footnote{Proceedings of the 9th
International Conference on Mobile Internet Security (MobiSec'25), Article No.
9, December 16-18, 2025, Sapporo, Japan. \ \textcopyright \ The copyright of
this paper remains with the author(s).}}




\ifdefined\anon

\author{Anonymous Author(s)}
\institute{Anonymous Institution(s)}

\else

\newcommand{\repeatthanks}{\textsuperscript{\thefootnote}}

\author{
  Sangryu Park \and
  Gihyuk Ko\thanks{Corresponding Authors} \and
  Homook Cho\repeatthanks
}

\institute{
  KAIST Cyber Security Research Center, Daejeon, Republic of Korea \\
  \email{\{sangryupark, gihyuk.ko, chmook79\}@kaist.ac.kr}
}

\fi

\maketitle



\begin{abstract}
Large Language Models (LLMs) show significant promise in automating software
vulnerability analysis, a critical task given the impact of security failure of
modern software systems. However, current approaches in using LLMs to automate
vulnerability analysis mostly rely on using online API-based LLM services,
requiring the user to disclose the source code in development. Moreover, they
predominantly frame the task as a binary classification (vulnerable or not
vulnerable), limiting potential practical utility. This paper addresses these
limitations by reformulating the problem as Software Vulnerability
Identification (SVI), where LLMs are asked to output the type of weakness in
Common Weakness Enumeration (CWE) IDs rather than simply indicating the presence
or absence of a vulnerability. We also tackle the reliance on large, API-based
LLMs by demonstrating that instruction-tuning smaller, locally deployable LLMs
can achieve superior identification performance. In our analysis,
instruct-tuning a local LLM showed better overall performance and cost trade-off
than online API-based LLMs. Our findings indicate that instruct-tuned local
models represent a more effective, secure, and practical approach for leveraging
LLMs in real-world vulnerability management workflows.

\end{abstract}

\keywords{Large Language Models (LLMs), Software Vulnerability, Vulnerability Detection, Vulnerability Identification, Local LLMs}


\section{Introduction}
\label{sec:intro}

The growing reliance on software across all areas of society has driven the
development of increasingly complex and high-performing software systems.
However, the risk of security failure grows alongside this advancement as larger
and more complex software tend to harbor more security vulnerabilities
\cite{murtaza16}. As a result, they often become attractive targets for
malicious attackers seeking to exploit them~\cite{telang07}. Reflecting this
trend, a number of published CVE (Common Vulnerabilities and Exposures) records
has more than tripled in the last decade, increasing from 59,487 (2005--2014) to
194,049 (2015--2024) records \cite{cverecords}. Yet, to this day, effectively
detecting and mitigating security vulnerabilities in software systems remains a
pervasive challenge.

To combat this, a numerous Static Application Security Testing (SAST)
\cite{lee06sa, gp15, oyetoyan18, li19, esposito24sast} methods have been
suggested over time. While they provide certain guarantee, they often suffer
from inherent limitations such as high false-positive rates, context blindness,
late-stage application, and lack of code-level precision~\cite{esposito24sast,
lenarduzzi23, guo23}. While Machine Learning (ML) and Deep Learning (DL) have
shown promise in overcoming some of these hurdles, the advent of Large Language
Models (LLMs) represents a potentially transformative shift in automated
vulnerability analysis due to their advanced code understanding capabilities
\cite{sheng25llmvul, omer23vul, guo24llmvul}.

Despite the potential of LLMs, current research on automating source code
vulnerability analysis with LLMs possess a few limitations. First, many proposed
works rely on online API-based LLM services such as GPT-4o and Gemini
\cite{tamberg25, yin24pro, yin24chatgpt, fu23chatgpt}, raising concerns about
privacy and intellectual property \cite{llmprivacy}. Additionally, many works
predominantly frame software vulnerability detection as a simple binary
classification problem -- determining if an isolated code snippet is simply
``vulnerable'' or ``not vulnerable''~\cite{sheng25llmvul, guo24llmvul}. This
simplification, often driven by benchmark dataset structures and ease of
evaluation, can cause limited actionability for remediation.

This paper addresses these critical gaps through following contributions.
Firstly, we focus on and reformulate the problem of \svifull, where we prompt
LLMs to predict not only whether a given piece of source code is vulnerable, but
also to identify the specific type of vulnerability. Secondly, we challenge the
reliance on larger, online API-based LLMs which entail significant cost and
privacy concerns. We investigate and demonstrate the effectiveness of
instruct-tuning smaller, locally deployable LLMs, showing they can achieve
superior performance on this specialized task. Lastly, we provide analysis
Common Weakness Enumeration (CWE) hierarchy as a framework for a structured
analysis of our model's identification capabilities across different weakness
categories and abstraction levels, moving beyond simplistic aggregate metrics.
By addressing these gaps, this work aims to advance LLM-based vulnerability
analysis towards more practical, secure, and effective solutions.

The remainder of this paper is organized as follows. Sections \ref{sec:related}
and \ref{sec:problem} provide background for our work by reviewing related work
and outlining the problem setting, respectively. Section \ref{sec:method}
presents our approach to instruction-tuning smaller, locally deployable LLMs for
\svifull. In Section \ref{sec:data}, we detail the datasets and models used for
evaluation. Sections \ref{sec:evaluation} and \ref{sec:discussion} report the
experimental results and offer corresponding discussions.

\section{Related Works}
\label{sec:related}

Over the recent years, many works have been suggested to effectively analyze
software vulnerability using Large Language Models (LLMs). In this section, we
review previously suggested LLM-based vulnerability detection methods from the
perspective of three areas: discriminative model-based, generative model-based,
and LLM agent-based.

\subsection{Discriminative Model-based Vulnerability Detection}

Early works in LLM-based vulnerability detection primarily used discriminative
models such as BERT, which are well-suited for classification and
detection tasks. For instance, Kim et al.~\cite{vuldebert} proposed VulDeBERT,
which detects code vulnerabilty by slicing the target program into code gadgets
and embedding them using BERT-based model. Since it relied on system calls to
detect vulnerabilities, VulDeBERT's detection scope is limited. Similarly, Fu et
al.~\cite{linevul} introduced LineVul, a CodeBERT-based line-level vulnerability
detection method. LineVul uses CodeBERT to generate vector representations of
code lines and checked if each line had any vulnerabilities. However, due to
parametrical constraints of BERT, LineVul is limited to handling inputs of up
to 512 tokens, making it unsuitable for analyzing longer code samples.

\subsection{Generative Model-based Vulnerability Detection}

Generative large language models were also actively utilized to detect code
vulnerabilities. For instance, Zhou et al.~\cite{examllm} examined vulnerability
detection performance of generative language models such as GPT-3.5 and GPT-4,
alongside with CodeBERT. For vulnerability detection, GPT-3.5 and GPT-4
outperformed state-of-the-art vulnerability detection model based on CodeBERT.
However, Zhou et al. pointed out that using API-based online LLM services such
as GPT-like models can cause data privacy and security issues since thet require
sending code to external servers.

To address the limitations of single-task models, Du et al.~\cite{vulllm}
introduced VulLLM that treats vulnerability detection as a multi-task problem.
Unlike other works, VulLLM was capable of localizing the potentially vulnerable
lines, detecting vulnerabilities and generating an explanation for it. However,
VulLLM had an overfitting problem where the accuracy of VulLLM significantly
drops when detecting vulnerability data from out of distribution.

\subsection{LLM Agent-based Vulnerability Detection}

Currently many of LLM works are based on LLM agents and also a number of
vulnerability detection research has been published. Purba et
al.~\cite{locagent} presents a multi-component LLM agent system for
vulnerability detection, where it uses checklist-guided LLM agents to analyze
source code. On the other hand, Guo et al.~\cite{repoaudit} proposed RepoAudit,
where autonomous LLM agents audit repository-level code to find vulnerabilities.
RepoAudit introduces dual validation to mitigate hallucinations and employs
demand-driven graph traversal to enhance scalability, achieving an average
analysis time of just 0.44 hours for 251,000 code files. Additionally, RepoAudit
claims that it can utilize LLM's implicit path discrimination without explicit
enumeration of sensitive paths.

While effective in certain scenarios, agent-based approaches can also face
notable limitations. Notably, Yarra~\cite{llmpatronous} points out that the
performance of LLM agent-based vulnerabilty detection is often dependent on the
underlying LLM, which may result in a high false negative rates. Furthermore,
Yarra~\cite{llmpatronous} claims that agent-based methods can be more
challenging to design, deploy, and maintain, as the architectures typically
requires multiple LLM inferences per vulnerability. This can introduce
significant computational overhead in terms of both cost and time. In
particular, the total processing time and cost tend to increase roughly in
proportion to the number of agents involved.

\section{Problem Setting}
\label{sec:problem}

In this section, we outline the problem setting under investigation, including
the assumptions, objectives, and settings relevant to our study.

\paragraph*{Local Analysis of Software Vulnerabilities}
In this work, we consider a situation where a user wishes to analyze software
vulnerabilities within a closed environment -- that is, without disclosing the
target source code. We argue that this is a highly plausible scenario, as
organizations -- particularly those developing innovative or sensitive
technologies -- may prefer to keep their products confidential until a full
public launch. Since it is a well-known fact that approximately 70\% of software
vulnerabilities stem from the defects in the development phase~\cite{omer23vul},
they have much incentives to deploy their own local LLM instead of having to
disclose their source codes online.

\paragraph*{\svifull}
In this work, we aim to address the problem of \emph{\svifull} for source code.
\svifull \ is similar to \svdfull \ in that SVD aims to detect \emph{whether or
not} a given source code is vulnerable, but it also aims to specify \emph{which
type of vulnerabilty} the given source code has. Naturally, while SVD can be
formulated as a binary classification task, SVI can be formulated as a
multi-class classification task.

Formally, we define the goal of the user as the following: Let $\{s_i\}$ denote
the set of source codes the user wishes to check and $\{v_i\}$ be the
corresponding vulnerability types (it can be benign). Let $C_{LLM}$ denote a
classifier identifying vulnerabilities based on LLM. The user's goal in our
setting will be to minimize $\sum_i \mathcal{L}(v_i,\bar{v}_i)$,
where $\bar{v}_i = C_{LLM}(s_i)$ and $\mathcal{L}(\cdot)$ is an appropriate loss
function.

\paragraph*{Common Weakness Enumeration (CWE)}
Common Weakness Enumeration (CWE)~\cite{mitre_cwe} is a structured list and
classification system for vulnerabilities that includes over 600 types of
vulnerabilities and provides taxonomy. In our problem setting, we assume that
the objective of the model is to identify CWEs. While Common Vulnerabilities
and Exposures (CVEs) are a more specific and practical list of weaknesses list
than CWEs, it is difficult to efficiently learn CVEs as they are often too
specific to the hardware and software configurations.

\section{Our Method}
\label{sec:method}

\begin{figure*}[t!]
    \centerline{
        \includegraphics[width=\textwidth, height=7cm]{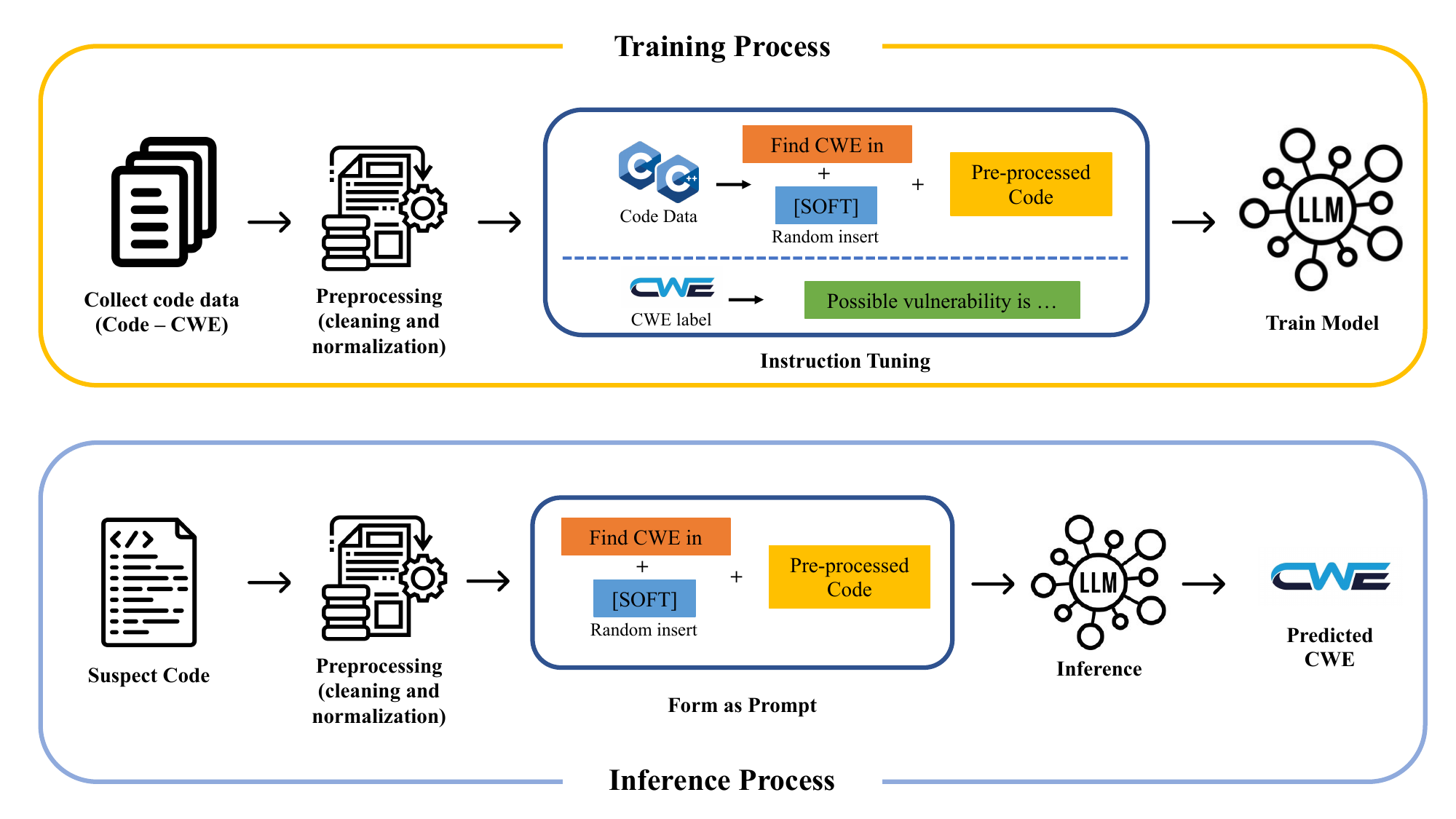}
    }
    \caption{Fine-tuning LLMs for Software Vulnerability Identification}
    \label{fig:architecture}
\end{figure*}

In this section, we detail our method of using LLMs for software vulnerability
identification. In our approach, we adopt an instruct-tuning~\cite{wei22flan}
method, where each data in a labeled dataset is first formatted in a predefined
textual format, and then used to fine-tune the backbone LLM. We argue that since
most decoder-based LLMs (which most modern generative language models are based
on) are trained for \emph{generative} tasks rather than \emph{discriminative}
tasks, our prompt-tuning method can be efficiently applied to enhance
vulnerability identification capabilities of LLMs.

An overview of our approach is presented in Fig.~\ref{fig:architecture}. Given a
dataset of [Code--CWE] pairs, we follow standard procedure to prepare code with
textual prompts for instruction tuning. We provide additional detail on the
collected dataset in Section~\ref{sec:data}.

\subsection{Preprocessing codes}
\label{ss:preprocessing}
We first remove unnecessary or noisy texts in the source code to prevent them
from disturbing training. For instance, special characters such as newline
(\texttt{\textbackslash n}) and tab (\texttt{\textbackslash t}) are removed.
Moreover, we normalize indentation by changing multiple tabs or spaces into a
single space per indent level. Lastly, as the comments in the source code can
be hinting towards the CWE IDs, they have been removed.

\subsection{Prompting}
\label{ss:prompting}
\begin{figure}[tb!]
    \centerline{\includegraphics[width=\textwidth]{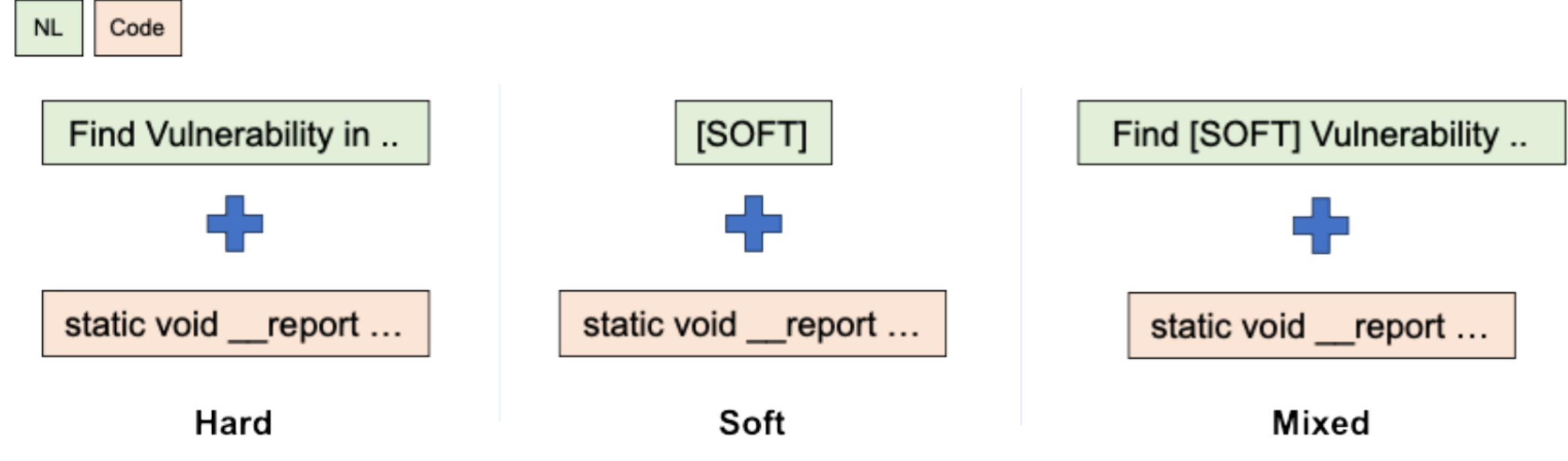}}
    \caption{Different prompting styles using [SOFT] token}
    \label{fig:prompt_style}
\end{figure}

Next, we combine preprocessed codes with textual queries to form prompts.
Following~\cite{prompt_rule}, we consider three different styles of prompting:
\texttt{hard}, \texttt{soft}, and \texttt{mixed prompting}. While textual query
is used as-is in \texttt{hard prompting}, the query is replaced with a specially
designated \textit{[SOFT]} token in \texttt{soft prompting}. In \texttt{mixed
prompting}, \textit{[SOFT]} tokens are randomly inserted in the textual queries.
Fig.\ref{fig:prompt_style} illustrates three different prompting styles.

We supplement each prompt with an expected response, which contains either the
description and/or the ID of the corresponding vulnerability (i.e., CWE), or a
message denoting that the code is not vulnerable. Note that there could be three
variants of the expected response: (1) CWE-ID with description, (2) CWE-ID only,
and (3) CWE description only. From our initial experiment, CWE-IDs seem to
confuse language models as the CWE IDs themselves are simply numeric, and do not
entail meaningful information. Therefore, we decided to use ``3) CWE description
only'' response.

\subsection{Instruction Tuning and Post-Processing}
\label{ss:tuning}
Lastly, we instruct-tune the model with prompt-augmented code data. During the
instruct-tuning process, the model is trained to extract features from input
code prompt and return CWE description as an answer. We apply a standard
instruction-tuning which corrects parameters based on how close the current
textual output is to the ground truth. Since outputs are purely textual, we
used BLEU score~\cite{bleu} to measure the discrepancies.

During the inference phase, a well-trained model may generate a CWE description
that differs slightly from the ground truth. To address this, we implemented a
post-processing mechanism that selects the closest matching CWE description from
a predefined set of possible descriptions. Our post-processing first selects the
most similar description based on the matching word counts. If there exist
multiple candidates with the same number of matching words, we select the
candidate with the highest BLEU score~\cite{bleu}.


\section{Data Collection and Model}
\label{sec:data}

Collecting sufficient, high-quality data samples and selecting a model
architecture well-suited to the task is a prerequisite for building a successful
AI system. This is especially true for \svi, as imbalances between vulnerable
and non-vulnerable code samples, or wrong choice of the model architecture can
significantly affect model performances. In this section, we describe the
process of constructing our dataset and explain our rationale for choosing
the base models.


\begin{table}[tb!]
    \setlength{\tabcolsep}{5pt}
    \renewcommand{\arraystretch}{1.3}
    \caption{Composition of the collected dataset.}
    \label{tab:dataset}
    \centering
    \begin{tabular}{c|c|c|c}
    \hline
    \textbf{CWE ID} & \textbf{Description} &\textbf{Rank} & \textbf{Counts} \\
    \hline \hline
    CWE-787 & Out of Bounds Write & 2 & 31,692 \\
    \hline
    CWE-125 & Out of Bounds Read & 6 & 23,161 \\
    \hline
    CWE-416 & Use After Free & 8 & 17,894 \\
    \hline
    CWE-20 & Improper Input Validation & 12 & 18,739 \\
    \hline
    \multirow{2}{*}{CWE-200} & Exposure of Sensitive Information & \multirow{2}{*}{17} & \multirow{2}{*}{10,890} \\
    & to an Unauthorized Actor & & \\
    \hline
    \multirow{2}{*}{CWE-119} & Improper Restriction of Operation within & \multirow{2}{*}{20} & \multirow{2}{*}{21,937} \\
    & Bounds of Memory Buffer & & \\
    \hline
    CWE-476 & Null Pointer Dereference & 21 & 15,121 \\
    \hline
    CWE-190 & Integer Overflow or Wraparound & 23 & 9,384 \\
    \hline
    \multirow{2}{*}{CWE-703} & Improper Check or Handling of & \multirow{2}{*}{-} & \multirow{2}{*}{19,910} \\
    & Exceptional Conditions & & \\
    \hline
    \multicolumn{3}{c|} {Benign} & 18,823 \\
    \hline \hline
    \multicolumn{3}{c|}{\textbf{Total}} & \textbf{187,551} \\
    \hline
    \end{tabular}
\end{table}

\subsection{Dataset}
\label{ss:data}
We design our model to identify vulnerabilities in code snippets written in
C/C++, as most datasets used in automatic vulnerability detection (or
identification) consist predominantly of C/C++ code~\cite{sheng25llmvul}. In our
evaluation, we collected vulnerable code data from BigVul~\cite{bigvul},
DiverseVul~\cite{diversevul}, and SVEN~\cite{sven}. To balance the dataset with
benign code samples, we collected code snippets from GNU Coreutils.

In selecting the target weaknesses (i.e., CWEs) for vulnerable code samples, we
consider two key factors: (1) sufficiency of the code samples for each CWE type,
and (2) the importance of the weakness in real-world scenarios. To ensure
effective training of LLMs, we prioritize CWEs with approximately 10,000 or more
code samples. Additionally, to reflect real-world threat patterns, we prioritize
those CWEs that appear in the CWE \textit{Top 25 Most Dangerous Software
Weaknesses} list~\cite{mitre}.

Table~\ref{tab:dataset} shows the composition of our dataset. Our dataset
consists of 187,551 code snippets with 10 different labels: 9 corresponding to
different CWE IDs and 1 representing benign code. We record the CWE IDs
alongside the description and rank (according to~\cite{mitre}) of each weakness.
For consistent evaluation, we randomly select 500 code snippets for each label
and form a test dataset consisting of total 5,000 code snippets. The remainder
(total of 182,551 code snippets) were used to train the backbone LLM model.

\subsection{Backbone Model}
\label{ss:model}
The architecture of Large Language Models (LLMs) varies by model, with each
offering distinct advantages depending on the underlying task. In the context of
\svi, encoder-based models such as CodeBERT~\cite{codebert} are particularly
effective due to their ability to learn rich semantic features from textual
input. These models can be optimized for comprehension tasks, enabling a deeper
understanding of source code and facilitating accurate vulnerability
identification.

In contrast, decoder-based models such as CodeLlama~\cite{codellama} are
typically specialized for generative tasks, including code generation and
translation. Although they are not primarily designed for feature extraction,
they can be trained with remembering substantial background knowledge. Moreover,
thanks to the use of prompting techniques, decoder-based models can be tuned to
perform numerous tasks beyond generation, including \svi.

In our experiments, we select CodeT5 (\codetfiveex)~\cite{codet5} as
our primary target of evaluation. We argue that since CodeT5 is based on the
encoder-decoder model T5~\cite{raffel20t5}, it can leverage the strengths of
both encoder- and decoder-based models. Specifically, the encoder is responsible
for extracting meaningful representations from the input code, while the decoder
utilizes this information, along with its pretrained knowledge, to generate
task-specific outputs.

To compare our method's efficacy, we also test on 5 different LLMs: CodeBERT
(\codebertex)~\cite{codebert}, GPT-3.5 (\gptthreeex)~\cite{gpt35}, GPT-4
(\gptfourex)~\cite{gpt4}, CodeLlama (\codellamaex)~\cite{codellama}, and Llama 3
(\llamathreeex)~\cite{llama3}. Note that GPT-3.5 and GPT-4 are only accessible
via paid API calls, and LLama 3 and CodeLlama are downloadable.


\section{Evaluation}
\label{sec:evaluation}

In this section, we present the experimental details and results used to assess
the effectiveness of our proposed method. Specifically, we design experiments to
address the following research questions (RQs):
\begin{list}{$\bullet$}{}
\item \textbf{RQ1. (Performance) } How effectively does the instruct-tuned local
LLM identify software vulnerabilities compared to online LLMs?
\item \textbf{RQ2. (Prompting) } Which instruct-tuning method is most effective
for identifying software vulnerabilities?
\item \textbf{RQ3. (Cost-efficiency) } What is the cost efficiency of
instruct-tuning a local LLM versus tuning an online LLM?
\end{list}

Throughout the remainder of this section, we present the results corresponding
to each of our research questions.

\paragraph*{Experimental setup}
For all experiments, we used a CodeT5 model (\texttt{codet5\-large}) trained
with our dataset, on a AMD EPYC 7543 CPU and 4 NVIDIA-A100 GPUs. We fixed the
maximum token length as 1,200, and set batch sizes as 4 for both training and
evaluation. We trained the model for maximum 10 epochs with early stopping
patience 3 to avoid overfitting.


\begin{table} [tb!]
    \setlength{\tabcolsep}{5pt}
    \renewcommand{\arraystretch}{1.3}
    \caption{Average performance of vulnerability identification by different LLMs. The results are averaged over five different random seeds to reduce the impact of variance in data sampling.}
    \label{tab:ex_rq11}
    \centering
    \resizebox{\textwidth}{!}{%
    \begin{tabular}{|c|c|c|c|c|c|c|}
    \hline
    \textbf{Model} & \textbf{CodeT5} &\textbf{CodeBERT} & \textbf{GPT-3.5} & \textbf{GPT-4} & \textbf{Code Llama} & \textbf{Llama3}\\
    \hline
    Accuracy & \textbf{81.71} & 73.36 & 10.14 & 11.06 & 9.62 & 10.76\\
    \hline
    Macro-F1 & \textbf{81.93} & 73.53 & 5.43 & 6.73 & 7.79 & 8.68\\
    \hline
    FNR & \textbf{0.09} & 0.11 & 4.32 & 13.66 & 14.69 & 1.92\\
    \hline
    FPR & \textbf{1.62} & 1.81 & 96.04 & 87.6 & 83.88 & 98.32 \\
    \hline
    \end{tabular}}
\end{table}

\begin{table} [tb!]
    \setlength{\tabcolsep}{3pt}
    \renewcommand{\arraystretch}{1.3}
    \caption{Label-wise performance of vulnerability identification. Results are based on a single fixed random seed (seed = 42).}
    \centering
    \label{tab:ex_rq1_1}
    \resizebox{\textwidth}{!}{%
    \begin{tabular}{|c|c|c|c|c|c|c|c|c|c|c|}
    \hline
    \textbf{CWE ID} & \textbf{119} & \textbf{125} & \textbf{190} & \textbf{20} & \textbf{200} & \textbf{416} & \textbf{476} & \textbf{703} & \textbf{787} & \textbf{No Vul} \\
    \hline
    \textbf{CodeT5} & 67.80 & 75.80 & 75.80 & \textbf{78.80} & \textbf{80.40} & \textbf{81.40} & \textbf{78.00} & \textbf{71.60} & \textbf{67.00} & \textbf{98.40}\\
    \hline
    \textbf{CodeBERT} & \textbf{74.60} & \textbf{76.60} & \textbf{80.60} & 77.00 & 71.00 & 81.00 & 74.20 & 67.60 & 48.80 & 98.20 \\
    \hline
    \textbf{GPT-3.5} & 1.20 & 4.20 & 2.20 & 0.40 & 0 & 6.80 & 66.40 & 6.80 & 11.20 & 3.00 \\
    \hline
    \textbf{GPT-4} & 17.00 & 14.80 & 5.80 & 8.60 & 0.40 & 10.80 & 47.60 & 1.80 & 1.00 & 12.40 \\
    \hline
    \textbf{Code Llama} & 3.60 & 12.80 & 11.00 & 16.60 & 6.00 & 17.00 & 2.00 & 16.00 & 15.20 & 15.60 \\
    \hline
    \textbf{Llama3} & 28.60 & 14.40 & 10.20 & 4.20 & 1.40 & 9.40 & 12.20 & 15.00 & 1.40 & 1.00 \\
    \hline
    \end{tabular}}
\end{table}

\subsection{RQ1: Vulnerability Identification Performance}
\label{ss:rq1}
The primary objective in automated vulnerability identification is to accurately
identify the specific vulnerabilities present in a given code snippet. To
address RQ1, we conducted experiments to evaluate the accuracy, Macro-F1
score, False Negative Rate (FNR) and False Positive Rate (FPR) of our model in
comparison with several baseline models including CodeBERT, GPT-3.5, GPT-4,
CodeLlama, and LLaMA 3, as we denoted in Section~\ref{ss:model}. Note that we
fine-tune CodeBERT on the same dataset as CodeT5, but instruction tuning is not
applied due to the model's limitations. For the rest models, we query each test
sample to obtain textual output, which are post-processed (see Section
\ref{ss:tuning}) to obtain matching CWE descriptions.

To ensure robustness, we repeated the experiment on 5 differently seeded
train-test data splits. In other words, CodeT5 and CodeBERT models were
trained on five different training splits, and tested on corresponding test
datasets. We record our average performance in Table~\ref{tab:ex_rq11}. Note
that \texttt{accuracy} is an average accuracy of 10 different labels (not
equivalent to \textit{vulnerable--not vulnerable} accuracy). According to
Table~\ref{tab:ex_rq11}, instruct-tuned CodeT5 model showed the best
performance of over 80\% accuracy, far exceeding other models. Larger LLMs,
believed to have better prediction performance, seemed to suffer in identifying
vulnerabilities in the underlying source code.

Table~\ref{tab:ex_rq1_1} presents per-label accuracy, measured on a single fixed
seed. Similar to above, we observed that instruct-tuned CodeT5 shows best or
second-best performance in all labels. While CodeBERT shows equally good
performance, GPT models seemed to suffer from predicting only one label. From
these results, we argue that instruct-tuned local LLMs can be as competent as,
or more effective in identifying software vulnerabilities than online, API-based
LLMs.


\begin{table} [tb!]
    \setlength{\tabcolsep}{3pt}
    \renewcommand{\arraystretch}{1.3}
    \caption{Impact of different prompting methods on the vulnerability identification performances.}
    \label{tab:ex_rq2}
    \centering
    \begin{tabular}{|c|c|c|c|c|}
    \hline
    \textbf{Model} & \textbf{\texttt{Mixed} prompt} &\textbf{No prompt} &
    \textbf{\texttt{Hard} prompt} & \textbf{\texttt{Soft} prompt} \\
    \hline
    Accuracy & \textbf{77.50} & 9.82 & 72.24 & 70.80 \\
    \hline
    Macro-F1 & \textbf{77.58} & 8.72 & 72.53 & 71.13 \\
    \hline
    \end{tabular}
\end{table}

\begin{table} [tb!]
    \setlength{\tabcolsep}{3pt}
    \renewcommand{\arraystretch}{1.3}
    \caption{Impact of textual prompt styles on the vulnerability identification perfomances.}
    \label{tab:ex_rq2_2}
    \centering
    \begin{tabular}{|c|c|c|c|c|}
    \hline
    \textbf{Prompt} & \textbf{Accuracy} & \textbf{Macro-F1} \\
    \hline
    \textbf{Simple} & \textbf{77.50} & \textbf{77.58} \\
    \hline
    C/C++ Specified & 74.02 & 74.22 \\
    \hline
    CodeT5-style & 74.94 & 75.10 \\
    \hline
    Descriptive & 73.16 & 73.42 \\
    \hline
    \end{tabular}
\end{table}

\subsection{RQ2: Optimal Prompting}
\label{ss:rq2}
As denoted in Section~\ref{ss:prompting}, we test various prompt tuning
strategies -- \texttt{hard}, \texttt{soft}, and \texttt{mixed} prompts -- to
guide the model behavior. \texttt{Hard} prompts provide clear task instructions,
helping to constrain the model's focus, while \texttt{soft} prompts offer
adaptability and can promote more diverse and creative outputs. In this
experiment, we evaluated how each prompt method impacts overall accuracy.

Table~\ref{tab:ex_rq2} presents the results. In our experiment, \texttt{mixed}
prompt method yielded the best performance. We suspect that \texttt{mixed}
prompting method introduces a degree of variability, allowing the model to treat
soft tokens as flexible context or controlled noise, which can enhance its
generalization during training.

We also test if different textual prompt styles can affect the detection
performance. We test with the following four variants of textual prompts to
instruct-tune and query the model:

\begin{list}{$\bullet$}{}
\item \textbf{Simple (default) ---} \texttt{Find CWE in:\ [CODE]}
\item \textbf{C/C++ Specified ---} \texttt{Examine the given C/C++ code snippet and detect vulnerabilities:\ [CODE]}
\item \textbf{CodeT5-style ---} \texttt{Defect:\ [CODE]}
\item \textbf{Descriptive ---} \texttt{Analyze the following code and identify potential security vulnerabilities in:\ [CODE]}
\end{list}

Table~\ref{tab:ex_rq2_2} summarizes the results. In our experiments, Simple
style of prompting was most effective. We hypothesize that, due to the
relatively smaller size of our instruct-tuned local model, descriptive prefixes
can disturb the model from correctly reasoning about the code. Note that longer
variants (C/C++ Specified and Descriptive) showed lower accuracies in comparison
to the shorter variants.

\subsection{RQ3: Cost Efficiency}
\label{ss:rq3}

One of the key challenges in training large language models (LLMs) is in the
high cost. Models with a large number of parameters and extensive training data
require GPUs with substantial VRAM and prolonged training times, which can lead
to implosion of expenses. This can further intensify when one has to rely on the
paid API-based queries.


\begin{table} [tb!]
    \setlength{\tabcolsep}{5pt}
    \renewcommand{\arraystretch}{1.5}
    \caption{Estimated cost of instruct-tuning LLMs on our vulnerability dataset. The cost has been estimated based on RunPod GPU service~\cite{runpod}.}
    \label{tab:ex_rq3}
    \centering
    \begin{threeparttable}
    \resizebox{\textwidth}{!}{%
    \begin{tabular}{|c|c|c||c|c|c|c|}
    \hline
    \textbf{Models} & \textbf{CodeT5} & \textbf{CodeBERT} & \textbf{GPT-3.5} & \textbf{GPT-4$^g$} & \textbf{CodeLlama} & \textbf{Llama3}\\
    \hline
    \textbf{Training} & \$431.20$^a$ & \$3.52$^r$ & \$9797.70 & \$30617.83 & \$2304.00$^a$ & \$2664.00$^a$ \\
    \hline
    \textbf{Inference} & \$0.22$^r$ & \$0.22$^r$ & \$0.04 & \$12.41 & \$0.66$^r$ & \$0.66$^r$ \\
    \hline
    \hline
    \textbf{Total} & \$431.42 & \$3.74 & \$9797.74 & \$30630.24 & \$2304.66 & \$2664.66 \\
    \hline
    \end{tabular}}
    \begin{minipage}{\textwidth}
    \begin{tablenotes}{
        \small
        \item[a] using 4$\times$ A100s (80GB ea.)
        \item[r] using 1$\times$ RTX3090
        \item[g] GPT-4 does not support fine-tuning service on OpenAI API, therefore calculation \\ is based on GPT-4.1}
    \end{tablenotes}
    \end{minipage}
    \end{threeparttable}
\end{table}

To simulate real-world deployment scenarios, we estimated the total cost
required to instruct-tune each model on our vulnerability dataset. These
estimates are based on actual training durations and the size of the training
data. Table~\ref{tab:ex_rq3} summarizes the estimated costs associated with
instruct-tuning various LLMs. For both training and inference, we referenced the
official pricing of API calls for hosted models, where applicable, or the cost
of GPU usage on the RunPod service based on the corresponding GPU (4$\times$
A100s for training, 1$\times$ RTX3090 for inference).

According to Table~\ref{tab:ex_rq3}, locally trained CodeT5 and CodeBERT shows
the lowest and second-lowest budget required. This can be expected as the sizes
of the local models are much smaller. Considering their high performance, we
argue that instruct-tuning local models are more cost-efficient than online
models in smaller, domain-specific tasks such as \svi.


\section{Discussion}
\label{sec:discussion}

In this section, we provide additional discussion on the effectiveness of our
approach: instruction-tuning local LLMs for \svi. Specifically, we first analyze
how CWE's hierarchy is correlated to the error rates. Next, we analyze
computational cost of instruction-tuning with each model. Lastly, we provide a
brief discussion on the function-level vs. whole-program vulnerability analysis.

\begin{figure*}[tb!]
    \centerline{
    \includegraphics[width=\textwidth, height=9cm]{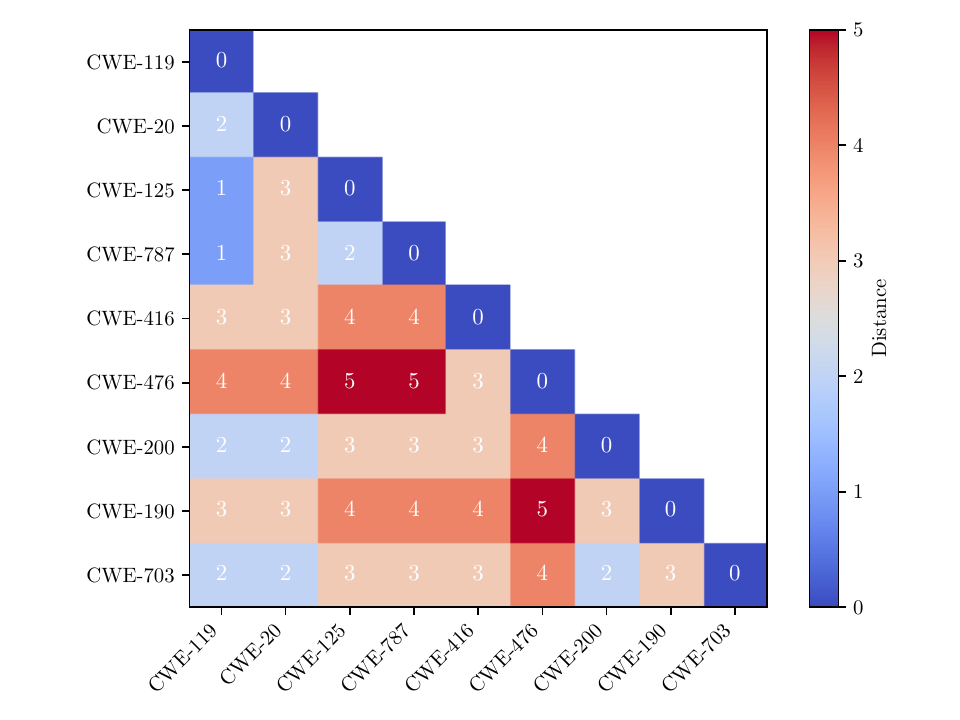}
    }
    \caption{Pairwise CWE Distance Matrix}
    \label{fig:dis_matrix}
\end{figure*}

\begin{figure*}[tb!]
    \centerline{
        \includegraphics[width=\textwidth]{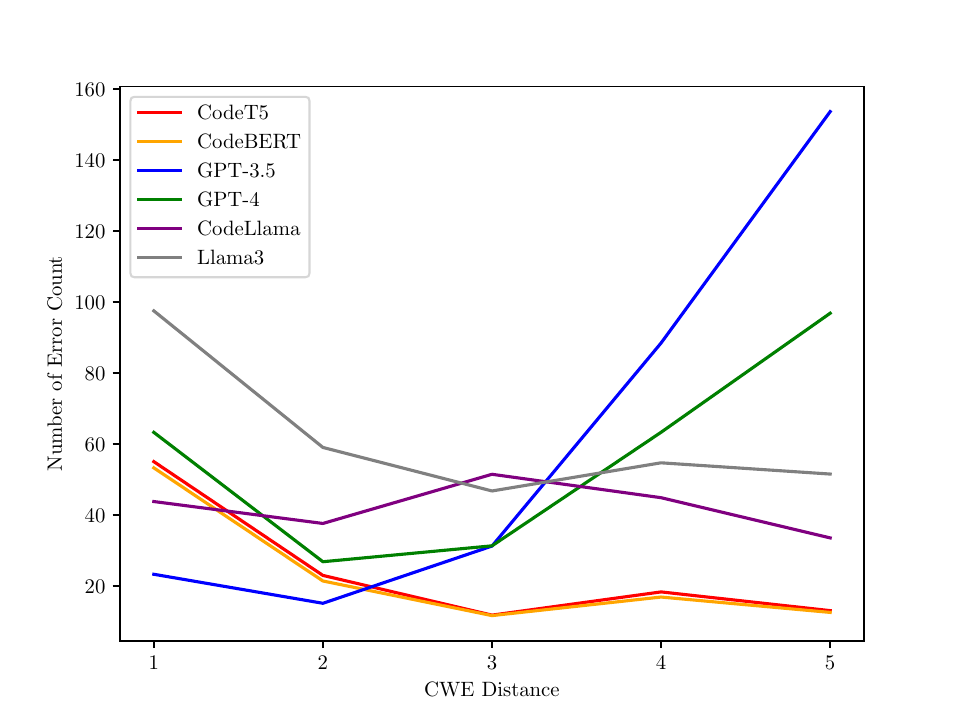}
    }
    \caption{Error Counts by Hierarchical CWE Distance}
    \label{fig:error_graph}
\end{figure*}

\subsection{Error by CWE Distance}
According to the Common Weakness Enumeration (CWE) taxonomy, each weakness has
an associated level of abstraction: Category, Class, Base, and Variant. These
abstraction enables CWEs to be connected on a hierarchical structure,
semantically relating two distant CWEs. For instance, CWE-125 (Out-of-Bounds
Read) is a Base CWE and falls under the broader CWE-119 (Improper Restriction of
Operations within the Bounds of a Memory Buffer), a Class-level CWE. In this
case, we can consider that CWE-125 and CWE-119 have a hierarchical distance of
1. Following this logic, we can map distance to every CWE pairs, as illustrated
in Figure~\ref{fig:dis_matrix}, where the hierarchical relationships are
quantified.

Based on this hierarchical distance measure between CWEs, we can analyze the
relationship between a model's prediction errors and the underlying CWE
distance. Naturally, it is expected that the model is likely to misclassify a
weakness as another that is semantically similar --- i.e., with lower CWE
distance. With this intuition, we plot a chart of error counts by CWE distances,
for all LLMs we have tested (Figure~\ref{fig:error_graph}).

Notably, GPT-3.5 and GPT-4 do not exhibit a meaningful correlation between CWE
distance and error frequency (note that in Figure~\ref{fig:dis_matrix}, there
exist more CWE pairs with distances 3 and 4 than distances 1 and 2). In
contrast, CodeT5 and CodeBERT demonstrate a clear expected pattern: the model
misclassifies more frequently when CWEs are more semantically similar. This
suggests that in many cases, these model successfully classify in a semantically
similar range of weaknesses. We argue that this shows a clear sign that in our
experiment, instruct-tuned models show a better performance in \svi \ than
online models. Additionally, these findings can imply that incorporating CWE
hierarchical information during training may further enhance model performance,
by reducing semantic misclassification among closely related weaknesses.


\begin{table} [tb!]
    \setlength{\tabcolsep}{5pt}
    \renewcommand{\arraystretch}{1.3}
    \caption{FLOPs amd MACs of models}
    \label{tab:ex_dis_1}
    \centering
    \resizebox{\textwidth}{!}{%
    \begin{threeparttable}
    \begin{tabular}{|c|c|c|c|c|c|}
    \hline
    \textbf{Model} & \textbf{CodeT5} &\textbf{CodeBERT} & \textbf{CodeLlama} & \textbf{Llama3} & \textbf{GPT$^*$} \\
    \hline
    FLOPs & \textbf{2.81E+12} & 2.76E+11 & 3.26E+14 & 5.79E+10 & $>$ 1.26E+24\\
    \hline
    MACs & \textbf{1.41E+12} & 1.38E+11 & 1.64E+14 & 2.89E+10 & $>$ 6.28E+23\\
    \hline
    \end{tabular}
    \begin{minipage}{\textwidth}
    \begin{tablenotes}[flushleft]
        \scriptsize
        \item[*] Since GPT-3.5 and GPT-4 model's architecture has not been opened, therefore calculation has been estimated with architecture of GPT-3
    \end{tablenotes}
    \end{minipage}
    \end{threeparttable}
    }
\end{table}

\subsection{Computational Cost}
Similar to in Section~\ref{ss:rq3}, we analyze computational cost required to
train and test each LLMs. Specifically, we use Floating Point Operations Per
Second (FLOPS)~\cite{flops} to estimate the computational cost of each
model. Based on FLOPS, we also estimated the total training time. Additionally,
we also estimate Multiply-Accumulate Operations (MACs) as a complementary metric
for assessing computational cost required to train and infer on a model. By
comparing FLOPS and MACs (as shown in Table \ref{tab:ex_dis_1}) alongside the
accuracy results of Table~\ref{tab:ex_rq11}, we argue that CodeBERT and CodeT5
offers the best trade-off between performance and computational cost for code
vulnerability detection.

\subsection{Function-level vs. Whole-program Vulnerability Analysis}
Our work is based on function-level vulnerability analysis, where the code
snippets contains a single function. While most of the previous works follow
this practice, there has been a growing need for a whole-program vulnerability
analysis. In this subsection, we aim to address the goods and limitations of
function-level vulnerability analysis.

While it is easier to collect, process, and train on function-level vulnerable
code snippets, we acknowledge that this approach is fundamentally limited by its
inability to capture out-of-function behaviors. Function-level analysis lacks
visibility into interprocedural context, such as data flow across functions,
global state changes, or interactions with external components. To
address this, some recent works have explored extending analysis to higher
levels of granularity, including interprocedural or whole-program analysis, to
capture more complex vulnerability patterns~\cite{song23, li24}.

Nevertheless, we argue that function-level vulnerability analysis can serve as a
crucial bridge between research-focused detection methods and real-world,
production-level security tools. Especially, under our problem setting where
users are limited to use smaller LLMs, connecting detection results from
function-level analysis to whole-program analysis would be a crucial next step.

\section{Conclusion}
\label{sec:conclusion}

The advent of generative AI tools and large language models (LLMs) has led our
society to rely on the complex software systems. In this fast-evolving
software landscape, ensuring the security and robustness of the software
systems has become more critical than ever. While LLM-assisted vulnerability
detection methods gives some promises in scalable detection and analysis of
underlying codes, they often rely on cloud-based services, raising concerns for
developers who prefer to keep their proprietary or sensitive code private during
development.

In this paper, we address a situation in which the users aim to automate the
process of LLM-based vulnerability identification locally, without uploading or
disclosing their source code in development. By utilizing a simple instruction-
tuning, we showed that even smaller encoder-decoder LLMs can be proven to be
both effective and cost-efficient for identifying software vulnerabilities. Our
findings show that with proper tuning and prompt strategies, smaller local
models can achieve competitive performance in vulnerability detection, while
preserving privacy and intellectual integrity.



\paragraph*{Acknowledgments}
\begin{small}
This work was supported by Institute of Information \& communications Technology
Planning \& Evaluation (IITP) grant funded by the Korea government(MSIT)
(RS-2023-00235509, Development of security monitoring technology based network
behavior against encrypted cyber threats in ICT convergence environment)
\end{small}

\bibliographystyle{plain}
\bibliography{aux/bibilography}


\end{document}